\documentstyle [12pt,epsfig]{article}
\begin{document}

\begin {center}
{\Large\bf The kappa in $J/\Psi \to K^+\pi ^- K^-\pi ^+$}

\vskip 0.3cm

D.V. Bugg

\vskip 0.2cm

\small

Queen Mary, University of London, London E1 4NS, UK
\vspace{0.2cm}

\normalsize

\end {center}
{\bf Abstract}

BES\,II data for $J/\Psi \to K^*(890)K\pi$ reveal
a strong $\kappa$ peak in the $K\pi$ S-wave near
threshold.
Both magnitude and phase are determined in slices of $K\pi$ mass by
interferences with strong $K_0(1430)$, $K_1(1270)$ and $K_1(1400)$
signals. The phase variation with mass agrees within errors
with LASS data for $K\pi$ elastic scattering. A combined fit is
presented to both BES and LASS data.
The fit uses a Breit-Wigner
amplitude with an $s$-dependent width containing an Adler zero. The
$\kappa$ pole is at $(760 \pm 20(stat) \pm 40(syst)) - i(420 \pm
45(stat) \pm 60(syst))$ MeV. The S-wave $I=0$ scattering length $a_0 =
0.23 \pm 0.04$ $m_\pi ^{-1}$ is close to the prediction $0.19 \pm 0.02$
$m_\pi ^{-1}$ of Chiral Perturbation Theory at $O(p^4)$.

\vspace{3mm}

\section {Introduction}
A possible $\kappa$ pole at low $K\pi$ mass is
controversial. Interest has been aroused by E791 data on $D^+ \to
K^-\pi ^+ \pi ^+$ [1], where there is evidence for a scalar resonance
of mass $M = 797 \pm 19 \pm 43$ MeV, width $\Gamma = 410 \pm 43 \pm 87$
MeV. However CLEO, with slightly lower statistics for $D^0 \to K^- \pi
^+ \pi ^0$, find no evidence for the $\kappa $ [2]. FOCUS data on $D^+
\to K^- \pi ^+ \mu ^+ \nu$ require $\bar K^{*0}$ interference with
either a broad spin zero resonance or a constant amplitude [3].

Wu has presented one solution for the $\kappa$ from the
BES data discussed here [4]. Komada has presented
a similar solution [5]. Both these analyses fit the $K\pi$ S-wave
using a conventional Breit-Wigner resonance plus a background
but do not describe this background.
Here a parametrisation based on Chiral Perturbation
Theory (ChPT) is used to fit the same data.
The Adler zero of ChPT leads to a strongly $s$-dependent width, which is
fitted to the data and replaces the background of Refs. [4] and [5].
This fit has the merit of also providing a good fit to LASS data on
$K\pi$ elastic scattering [6].
Oller has shown [7] that the E791 data of
Ref. [1] may also be fitted in a similar fashion consistently with LASS
data.

Theoretical opinion on the nature of the $\kappa$ is divided.
Scadron [9] argued in favour of an SU(3) nonet made up of
$\sigma$, $\kappa (800)$, $f_0(980)$ and $a_0(980)$.
The J\" ulich group of Lohse et al. [9] fitted LASS data
with $t$-channel exchanges, but
without requiring any $\kappa$ pole.
Further fits to LASS data  by Oller et al., based on a
unitarisation of chiral perturbation theory, find a $\kappa$ pole at
M$= 770 - i341$ MeV in Ref. [10] and at $708 - i305$ MeV in Ref. [11]; a
later paper of Pel\' aez and G\' omez Nicola quotes $(754 \pm 22) -
i(230 \pm 27)$ MeV [12]. Schechter et al. [13] also argue in favour of a
scalar nonet made of $\sigma$, $\kappa$, $f_0(980)$ and $a_0(980)$.
Van Beveren and Rupp have fitted the $K\pi$ S-wave and conclude
there is a $\kappa$ pole at $714-i228$ MeV [14]. However, Cherry and
Pennington [15] assert from LASS data that `There is no
$\kappa (900)$', though `data do not rule out a very low mass $\kappa$
below 825 MeV'. In view of the variety of conclusions, guidance from
experiment is obviously needed.

\section {Experimental Details and Data Selection}
The data selection is conventional and will be outlined only
briefly.
The data are from 58M $J/\Psi$ hadronic decays in the
upgraded BES II detector [16,17].
Charged particles are detected in a vertex chamber and
the Main Drift Chamber; these lie inside a solenoidal
magnet providing a uniform field of 0.5T.
A shower counter is used here
purely as a veto for photons; it is made of 12 radiation lengths
of lead sheets, interleaved with streamer chambers.
Kaons, pions and protons are identified up to 700 MeV/c by a
time-of-flight (TOF) array immediately outside the Main Drift
Chamber.
The $\sigma$ of the TOF measurement is 180 ps.
Further separation is obtained using $dE/dx$ in the Main Drift Chamber.
The vertex is required to lie within 2 cm of the beam axis and within
20 cm of the centre of the interaction region.
All particles are required to lie well
within the acceptance of the detector, with charged tracks having polar
angles $\theta$ of $|\cos \theta | < 0.80$.

The slowest two particles always have total energies $<800$ MeV.
For such energies, kaons and pions differ in momentum by at least 15\%.
The kinematic separation between $K$ and $\pi$ in a fit to
$K^+\pi ^-K^-\pi ^+$ is excellent.
In addition, both TOF and $dE/dx$ are used to calculate $\chi ^2$
probabilities for kaons or pions.
Monte Carlo studies show that the highest combined
probability selects the correct $K^+\pi ^-K^-\pi ^+$ combination for
almost all events, using a cut at $\chi ^2 = 40$.
An overall probability is required for
$K^+K^-\pi ^+\pi ^-$ higher than for $\pi ^+\pi ^- \pi ^+\pi ^-$ or
$K^+K^-K^+K^-$ or $K^\pm \pi ^\mp \pi ^+\pi ^-$. Any $K\pi \pi \pi$
combination with $M(\pi ^+ \pi ^-)$ in the interval $497 \pm 25$ MeV is
rejected if $r_{xy} > 8$ mm, where $r_{xy}$ is the distance from the
beam axis to the $\pi ^+\pi ^-$ vertex.
This procedure avoids producing a deep cut in genuine $K^+K^-\pi
^+\pi ^-$ events near the $K^0$ mass in $\pi ^+\pi ^-$.
The Monte Carlo
simulation estimates 215 background events from this source, widely
dispersed in $K\pi$ mass. Background from $\phi (1020)\pi ^+\pi ^-$ is
eliminated for $|M(K^+K^-) - M(\phi )| < 20$ MeV. Eventually, there are
77925 selected events. The final state $K^+\pi ^- K^- \pi ^+$ has a
large branching fraction $\sim 7 \times 10^{-3}$. Small surviving
backgrounds arise from many channels with an extra photon or $\pi ^0$;
Monte Carlo simulations estimate a surviving background of $3.2\%$,
which is fitted as $KK\pi \pi$ phase space.

\begin{figure}[htbp]
\begin{center}
\epsfig{file=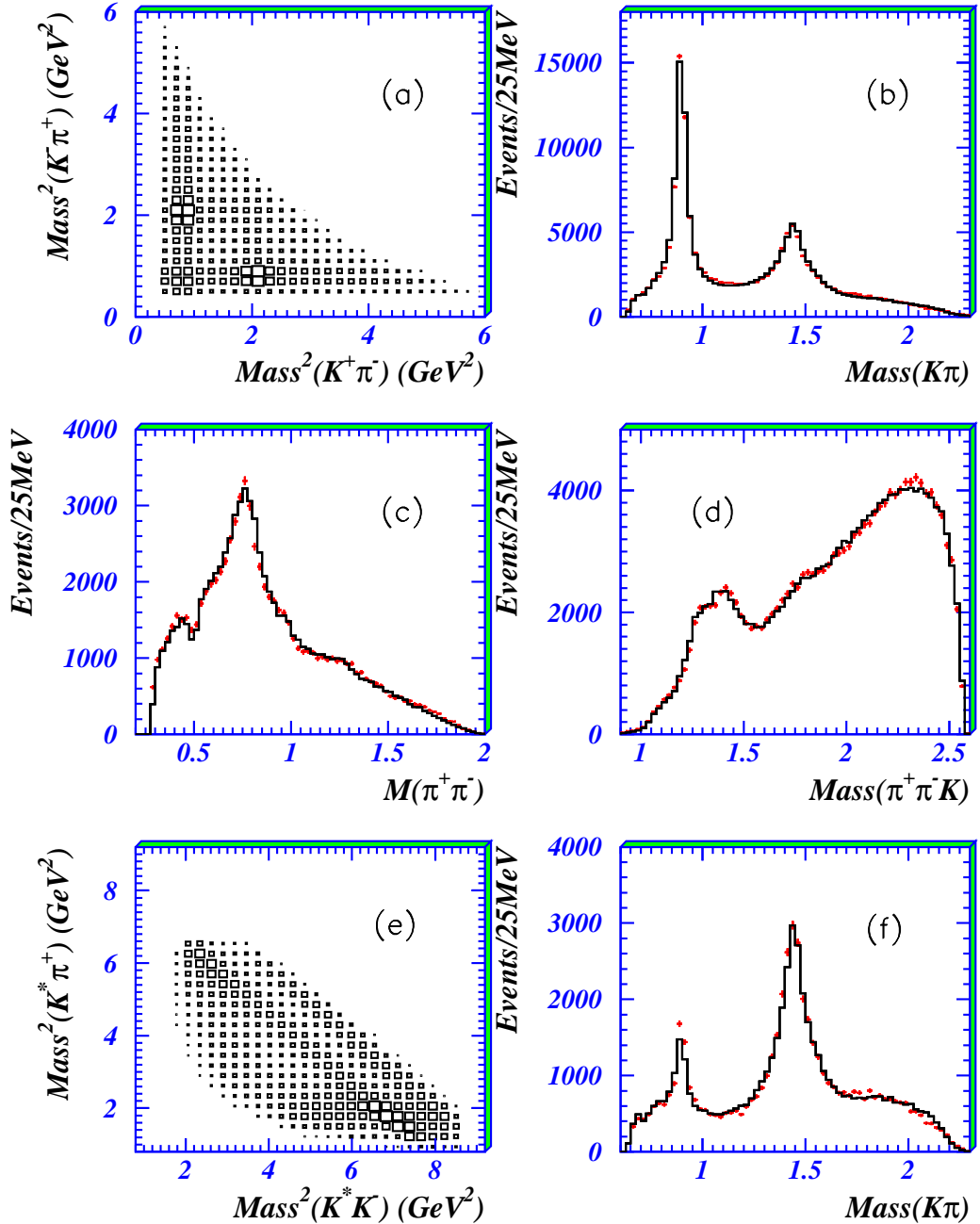,width=14.0cm}\
\end{center}
\caption[]{(a) The scatter plot of $M(K^+\pi ^-)$ against $M(K^-\pi
^+)$.
(b) Projection on to $K^\pm \pi ^\mp$ mass; the histogram shows
the fit.
(c) and (d) Projections on to $\pi \pi $ and $K\pi \pi $ mass.
(e) The Dalitz plot for
events where one $K^\pm \pi ^\mp$ combination is in the mass range
$892 \pm 100$ MeV.
(f) Mass projection of the second $K^\mp \pi ^\pm$ pair for
the same selection as (e).}
\end{figure}

\section {Features of the Data}
Fig. 1(a) shows the scatter plot of
$(K^+\pi ^-)(K^-\pi ^+)$ combinations.
There are obvious bands due to
$K^*(890)$ and around 1430 MeV, where there are three known
$K^*$ resonances.
Fig. 1(b) shows corresponding peaks in the $K\pi$ mass
projection and histograms show the fit described below.
Most of the peak at 1430 MeV is fitted as $K_0(1430)$,
with only a small contribution from $K_2(1430)$.
The $\rho (770)$ is visible in (c). In (d)
a strong peak appears due to overlapping $K_1(1270)$ and $K_1(1400)$;
there is also a weak $K_2(1770)$ peak.

Evidence for the $\kappa$ appears in the channel
$J/\Psi \to K^*(890)\kappa$, $\kappa \to (K\pi)_S$, where
$S$ denotes the S-wave.
To illustrate this, one $K^\pm \pi ^\mp$ pair (say particles 1 and 2)
is selected in the mass range $892 \pm  100$ MeV.
Then Fig. 1(f) shows the mass projection of the other $K^\mp
\pi ^\pm$ pair (particles 3 and 4). Peaks due to $K^*(890)$ and
$K_0(1430)/K_2(1430)$ are visible, but there is also a broad
$K\pi$ enhancement under the narrow $K^*(890)$.
Fig. 1(e) shows the Dalitz plot of $K^*K\pi$ combinations with this
data selection.

Fig. 2 shows the low mass $\kappa$ peak more clearly by
dividing the $K\pi$ mass projection  of Fig. 1(f) by
$K\pi$ phase space.
A definite peak becomes visible in the mass range
from threshold to $\sim 750$ MeV.
Analysis given below shows
that the $\kappa$ peak is quite broad, so one should ignore the
statistical fluctuation in the bin at 710 MeV.

\begin{figure}[htbp]
\begin{center}
\epsfig{file=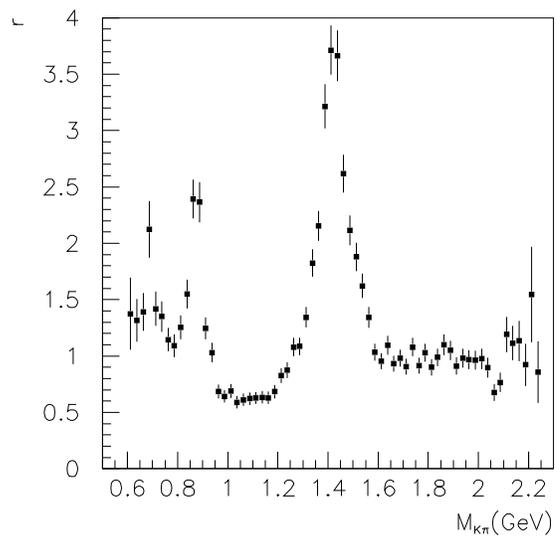,width=9.0cm}\
\end{center}
\vskip -1cm
\caption[]{The $K\pi$ mass projection of Fig. 1(f) divided by
$K\pi$ phase space, in bins of 20 MeV.}
\end{figure}

The $K^*(890)$ peak in Fig. 2 does not originate from
$K^*K^*$, which makes a negligible
contribution. Instead it originates as follows.
Underneath the $K^*(890)$ signal of Fig. 1(b) is some
$\kappa$ signal, which is selected by the cut of $\pm 100$
MeV around the $K^*(890)$; these $\kappa$ events are
accompanied by $K^*$ which create the narrow 890 MeV peak
in Fig. 2. This is the charge conjugate process.
Both combinations are included coherently into the
amplitude analysis.

In Fig. 1(e), the strong diagonal band
across the centre of the Dalitz plot is due mostly to $K_0(1430)$.
The weaker diagonal band at the upper right-hand edge
has a broad component due to the $\kappa$ and also a
narrow component from surviving $K^*(890)$.
The horizontal band at the bottom of the plot is due to
$K_1(1270)$ and $K_1(1400) \to K^*(890)\pi$.

Across Fig. 1(e),
there is a substantial ($\sim 15\%$) physics background arising from
$a_2\rho$, $a_0\rho$ and $K_1 \to K\rho$; the total $\rho$ contribution
is tightly controlled by fitting the magnitude of the $\rho$ peak in
Fig. 1(c). These channels produce the rather uniform background
visible across the scatter plot of Fig. 1(a).

\section {Amplitude Analysis}
The amplitude analysis fits events over all of 4-body phase space
to channels listed in Table 1.
This fit follows the standard isobar model, where each amplitude
is assigned a complex coupling constant.
Data are fitted by the maximum likelihood method to
relativistic tensor amplitudes of Ref. [18], using 600K Monte Carlo
events.
Angular momenta $L$ up to 2 in the production process are needed.
Standard Blatt-Weisskopf centrifugal barrier factors
given in Ref. [18] are included using a radius of interaction of 0.8
fm, though results are insensitive to this value.

\begin {table}[htp]
\begin {center}
\begin {tabular} {|lc|}
\hline
~~~Channel~~~~ & ~~~~~Percentage of events~~~~~ \\\hline
$K^*(890)K_0(1430)$ & $30.7 \pm 3.5$ \\
$K^*(890)\kappa   $ & $18.8 \pm 1.5$  \\
$K^*(890)K_0(1950)$ & $2.8 \pm 0.6$  \\
$K^*(890)K_2(1430)$ &  $7.6 \pm 0.5$  \\
$K_0(1430)\kappa  $ &  $6.8 \pm 1.8$  \\
$K_0(1430)K_0(1430)$& $4.6 \pm 0.7$    \\
$K_2(1430)K_0(1430)$& $2.2 \pm 0.2$   \\
$K_1(1400)K$        & $7.9 \pm 0.6$ \\
$K_1(1270)K$        & $12.3 \pm 1.1$ \\
$K_2(1750)K$        & $1.0 \pm 0.3$ \\
$a_0(980)\rho (770)$ & $1.6 \pm 0.5$ \\
$a_2(1320)\rho (770)$& $1.1 \pm 0.5$ \\
$a_2(1700)\rho (770)$& $3.3 \pm 0.9$  \\
$a_2(1990)\rho (770)$& $1.9 \pm 0.6$  \\
$a_2(2270)\rho (770)$& $3.4 \pm 0.3$  \\
$\phi (1680) f_2(1270)$ & $1.3 \pm 0.4$ \\
$\phi (1680) f_0(980)$ & $0.8 \pm 0.2$ \\\hline
\end {tabular}
\caption {Percentage of events fitted to every channel}
\end {center}
\end {table}

Table 1 shows the fraction of events fitted to each channel,
omitting interferences between channels, but keeping those
between different $L$ values within one channel;
percentages of events do not add up to exactly 100\% because of these
interferences.
Any amplitude which improves log likelihood by $<30$
is omitted; this eliminates
possible channels contributing $<1\%$
of events, e.g. unknown $K^*$ above 2 GeV.
The dominant channels are $K^*(890)K_0(1430)$,
$K^*(890)\kappa$, $K_1(1270)K$ and $K_1(1400)K$; all other channels have
little overlap with the essential $K^*(890)\kappa$ signal and very
little influence on parameters fitted to the $\kappa$. Most resonances
are fitted with masses and widths of the Particle Data Group [19].
However, $K_1(1270)$ optimises at $M = 1248 \pm 15$ MeV, $\Gamma = 157
\pm 35$ MeV, rather wider than the PDG value $\Gamma = 90 \pm 20$ MeV.
If PDG values are used instead, parameters for the $\kappa$ change
little, but the overall log likelihood of the fit gets worse.

There is a large $K_0(1430)$ signal in
the $K\pi$ S-wave as well as the $\kappa$.
It interferes strongly with the $\kappa$, and it is important
to separate $K_0(1430)$ from
$K_2(1430)$ and possible $K^*(1410)$ ($J^P = 1^-$).
A vital technical feature of the analysis is that
decays of the $K^*(890)$ are fitted; that was not done
in Refs. [4] and [5].
Angular correlations involve 5 angles between (i) the $K^*(890)$
decay, (ii) the production angle for the $K^*(890)$ and (iii)
decays of components of the 1430 MeV peak;
these correlations provide a secure separation of different $J^P$,
and demonstrate that most of the 1430 MeV peak is due to
$[K^*(890)K_0(1430)]_{L=0}$.
There is no evidence for any significant $K^*(1410)$. If
fitted freely, it contributes only 0.1\% of all events.
This is not surprising, in view of its $(6.6 \pm 1.3)\%$
branching ratio to $K\pi$ [19].

It is important to ensure that the $\kappa$
is not a `reflection' due to decays of
$K_1(1400)$ and $K_1(1270)$; both populate the low mass $K\pi$ region.
This possibility is eliminated by fitting $K_1(1270)$,
$K_1(1400)$ and $K_2(1770)$ to the peaks in
Fig. 1(d);
D and S-wave decays of both $K_1$ are included.
If the $\kappa$ signal were a `reflection' from these
sources, there would be little change to the fit if
the $\kappa$ is omitted. In fact
log likelihood changes by $>1000$, a very large amount.
A further detail is that $K^*(1400) (J^P=1^-) \to K^*(890)\pi$
has been tried in the fit.
A free fit contributes $<0.8\%$ of all events, and falls
below the cut-off for significant contributions to log likelihood.

\subsection {Formulae}
The motivation for the formula used to fit the $\kappa$
will now be discussed.
The essential point is to account for the peak in BES data
and its absence in elastic scattering.
This difference originates from the Adler zero in elastic
scattering, which largely cancels the pole.
If strong interactions
are chirally symmetric, massless pions of zero momentum have zero
scattering amplitude.
Breaking of chiral symmetry gives the pion a non-zero mass and Weinberg
[20] proposed that `soft' pions of low momentum $p$ have a matrix
element approximately linear in $p^2$ and $m^2_\pi$.
There is then a zero in the scattering amplitude for real pions at
$s \simeq m_K^2 - 0.5m^2_\pi$; this Adler zero is a central feature
of Chiral Perturbation Theory.
It is included here explicitly into an $s$-dependent width
$\Gamma (s)$ for the kappa.

The elastic scattering amplitude is written as [21]:
\begin {eqnarray} f_{el} &=& \frac {M_1\Gamma (s)}{M_1^2 - s -
iM_1\Gamma (s)} \\
\Gamma &=& \Gamma _0 (s - s_A)\exp (-\alpha \sqrt {s})\rho (s),
\end {eqnarray}
where $\rho (s)$ is $K\pi$ phase space $2k/\sqrt {s}$,
and $k$ is momentum in the $K\pi$ rest frame;
$\Gamma _0$ and $\alpha$ are constants.
This is the simplest realistic formula containing the Adler zero.
In the production  process, pions are `hard' because of the large
momentum transfer  between $J/\Psi$ and the final state.
BES data are then fitted with a complex coupling constant $G_{J/\Psi
}$:
\begin {equation} f_{prod} = \frac {G_{J/\Psi}}
{M_1^2 - s - iM_1\Gamma (s)}.
\end {equation}
Since the Adler zero is a feature of the full $K\pi$ S-wave amplitude,
the  $K_0(1430)$ is fitted with the Flatt\' e formula
\begin {equation} \Gamma (s) = \frac {s - s_A}{M_2 - s_A}
[g_1\rho _{K\pi }(s) + g_2\rho _{K\eta '}(s)];
\end {equation}
however, the Adler zero
plays only a small role for $K_0(1430)$ in practice.

For elastic scattering, unitarity is satisfied by adding the phases
of $K_0(1430)$ and $\kappa$; this is
the Dalitz-Tuan prescription [22].
For $J/\Psi$ decays there are hundreds of open channels, so
unitarity no longer makes any effective constraint; in this case,
the standard procedure of adding the amplitudes is used.
Note that it is being assumed that the $\kappa$ peak may be fitted
as a resonance, i.e. a pole. If there were a substantial non-resonant
background this would not be true. At the end of the analysis,
the possible requirement for background will be examined, but
no significant evidence for such a background emerges.
The stability of the $\kappa$ pole will be examined by trying a
variety of other assumptions for the $s$-dependence of both resonances.

Parameters of $K_0(1430)$ are constrained to fit
the peaks in both BES and LASS data [6]: BES data determine the
mass and width best, but LASS data determine better the ratio
$g_2/g_1$. Parameters for the $K_0(1430)$ optimise at $M _2= 1.535 \pm
15(stat) \pm 10(syst)$ GeV, $g_1 = 0.361 \pm 15 \pm 20$ GeV, $g_2/g_1 =
1.0 ^{+0.3}_{-0.2}$. The pole lies at M$_2 = (1433 \pm 30 \pm 10)- i
(181 \pm 10 \pm 12)$ MeV. Note that the full width of a conventional
Breit-Wigner resonance is twice the imaginary part of M$_2$;
the width quoted by the PDG is $294 \pm 23$ MeV.

The $\kappa$ line-shape is highly distorted
in elastic scattering by the Adler zero.
The phase shift must follow the
unitarity relation $f_{el} = \sin \delta e^{i\delta}/k$
up to the inelastic threshold.
The phase shift required by LASS data rises slowly
from threshold and passes  $90^\circ$ only close to $K_0(1430)$.
This requires a value of $M_1$ in equn. (1) well above 1430 MeV,
despite the pole near threshold. This unusual feature is
discussed by Zheng et al. [23], who propose an $s$-dependent
width similar to that adopted here.
The value of $M_1$ may be fitted anywhere in the range
2.4 to 4.0 GeV if $\Gamma _0$ is re-optimised.
The fit given here uses $M_1 = 3.3$ GeV, $\Gamma _0 =
24.53$ GeV, $\alpha = 0.4$ GeV$^{-1}$, though these are highly
correlated.
Extrapolating the amplitude off the real $s$-axis, the pole
is at M$_1 = (760 \pm 20(stat) \pm 40(syst)) - i(420 \pm
45(stat) \pm 60(syst))$ MeV.
The corresponding full-width is large: $840$ MeV;
systematic errors will be discussed below. The
width is much larger than the value $410 \pm 43 \pm 87$ MeV found by
E791 [1]; they did not include the Adler zero, but they
did include an interfering flat background over the Dalitz plot.

\begin{figure}[htbp]
\begin{center}
\centerline{\hspace{0.2cm}
\epsfig{file=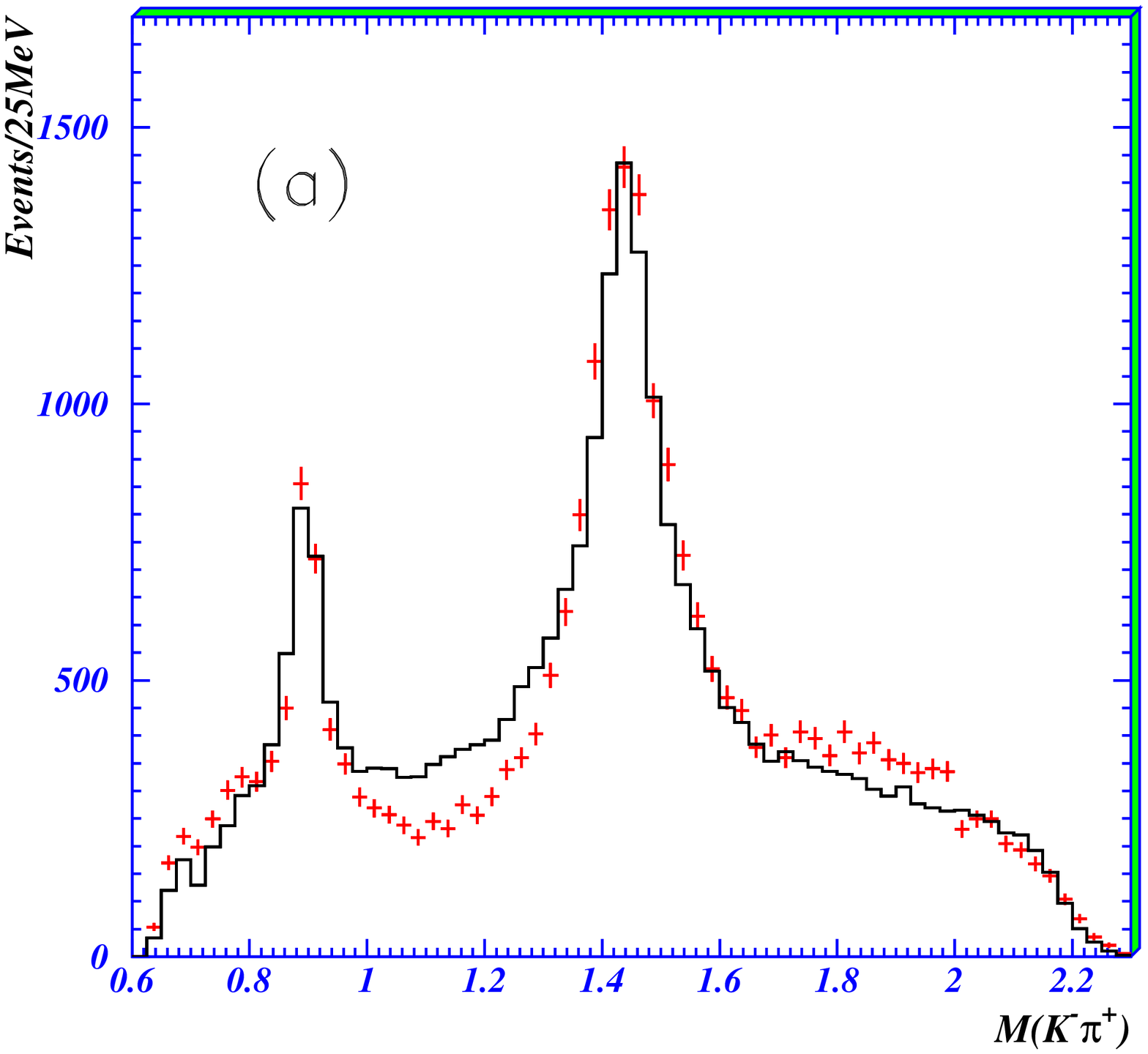,width=7.1cm}
\epsfig{file=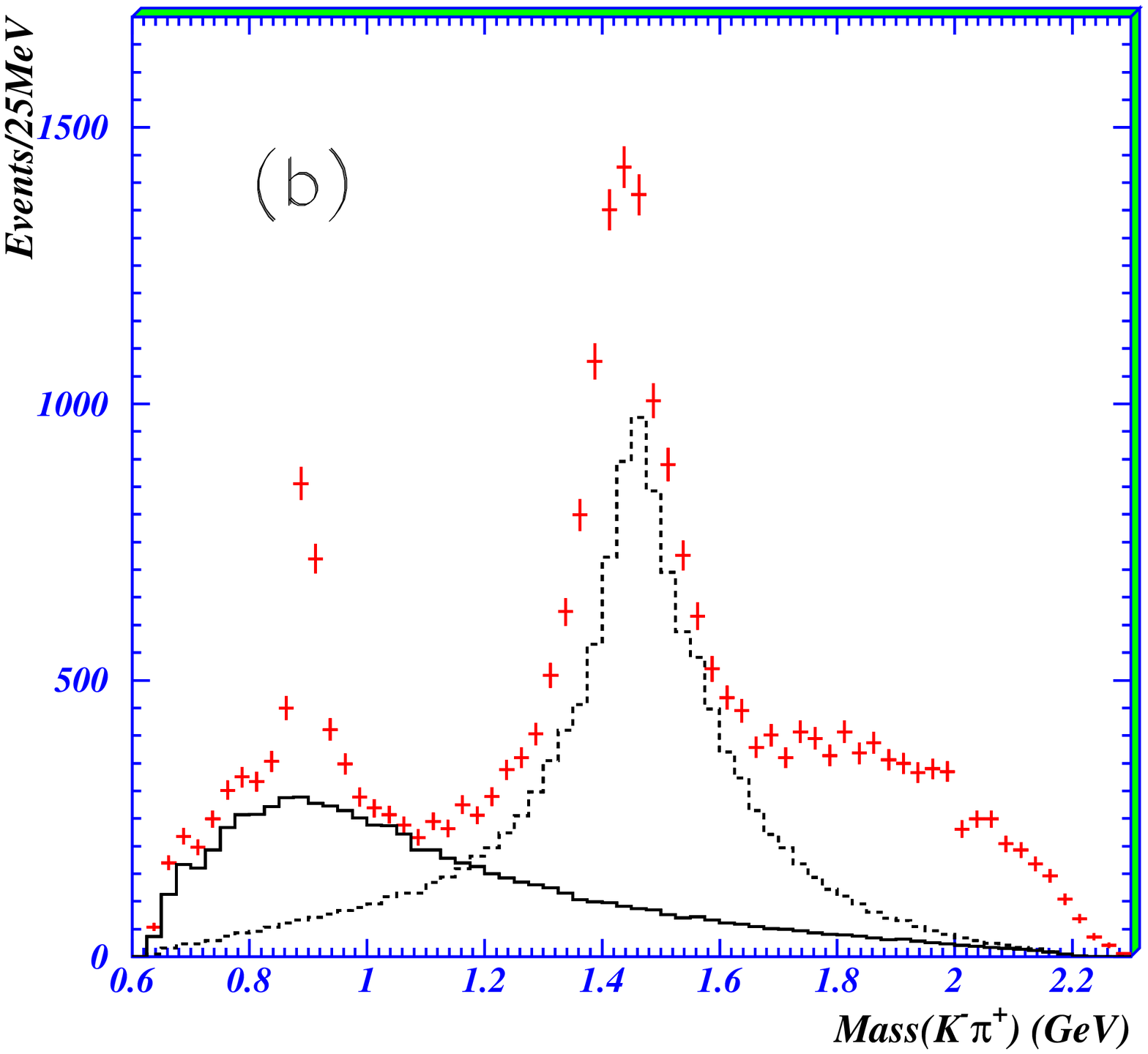,width=7.1cm}}
\vspace{0.1cm}
\centerline{\hspace{0.3cm}}
\epsfig{file=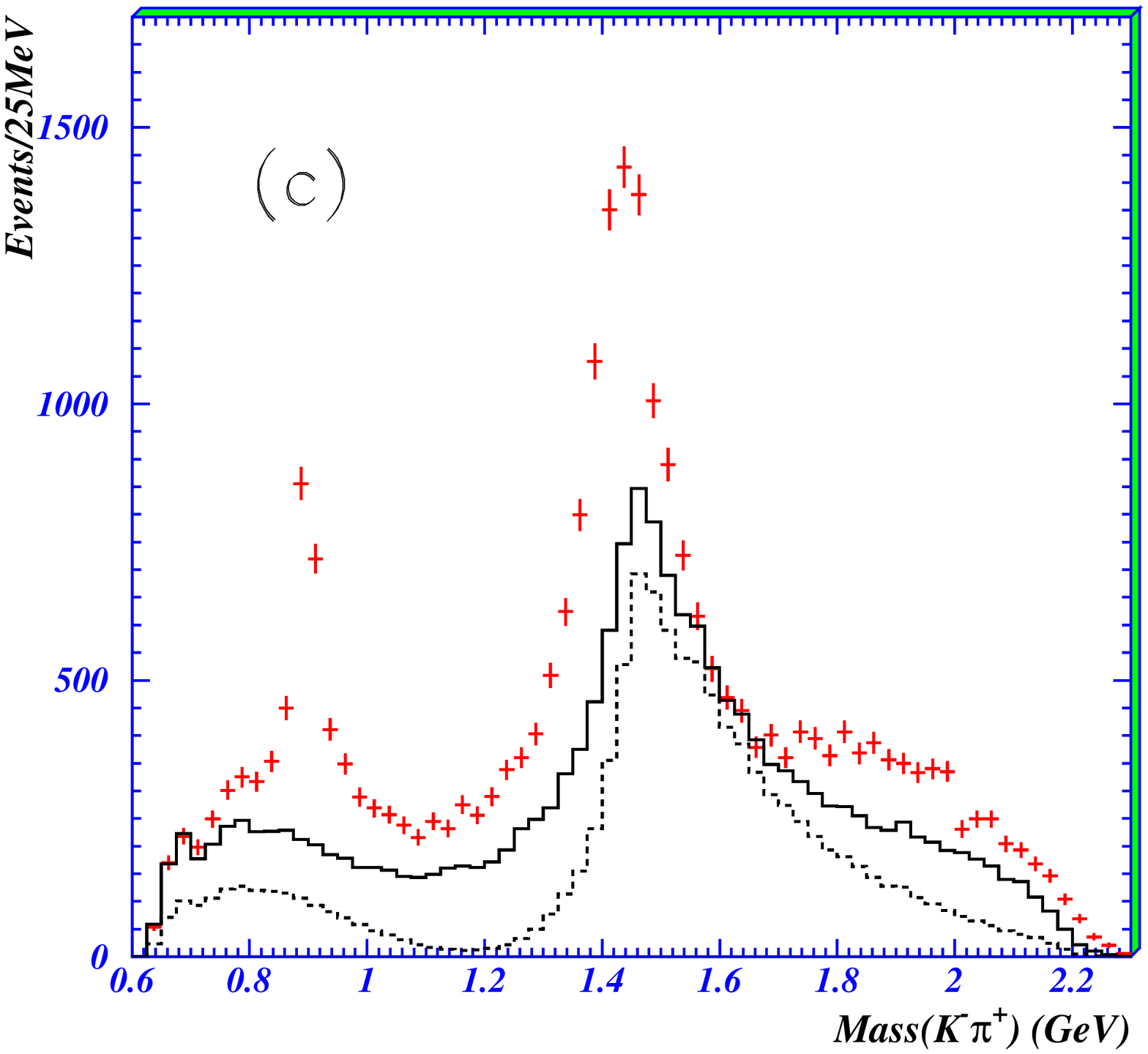,width=7.1cm}
\vspace{-0.1cm}
  \caption[]{(a) The poor fit when the channel $K^*(890)\kappa$ is
removed; (b) individual contributions from $\kappa$ (full histogram)
and $K_0(1430)$ (dashed); (c) the coherent
sum of $\kappa$ + $K_0(1430)$ (dashed histogram) and the coherent sum
$\kappa$ + $K_0(1430)+ K_1(1270) + K_1(1400)$ (full).}
\end {center}
\end{figure}

Removing $K^*(890)\kappa$ gives the fit shown by the histogram of Fig.
3(a); this is obviously unsatisfactory.
Using the scattering length formula of LASS gives a very
similar fit, because there is no $\kappa $ peak at low mass.
For the final fit, Figs. 3(b) and
(c) show contributions to the $K\pi$ mass projection with the
data selection of Figs. 1(e) and (f).
Fig. 3(b) shows intensities of $\kappa$ (full histogram)
and $K_0(1430)$ (dashed). On Fig. 3(c), the dashed histogram shows the
coherent sum of $\kappa$ and $K_0(1430)$; there is strong destructive
interference between them, simulating a low mass peak of width $\sim
400$ MeV.
The full histogram of Fig. 3(c) shows the coherent sum of
$\kappa$, $K_0(1430)$, $K_1(1270)$ and $K_1(1400)$. The difference
between data and the full histogram arises from (i) further
contributions from $K^0(1430)K^0(1430)$ and $K^0(1430)\kappa$, (ii)
interferences with other components, mostly $K_1K$ with
$K_1 \to K\rho$.

\subsection {The width of the $\kappa$}
In view of the width of the low mass $\kappa$ peak $\sim 400 $ MeV
and the similar width fitted by E791, an attempt has been made to
force the global fit to this width.
It produces a distinctly different solution with large changes in the
phases fitted to $K_1(1270)$, $K_1(1400)$ and $K_0(1430)$.
Such a solution is  metastable over a limited range of parameters.
However, it readily collapses to the fit reported here, with an
improvement in log likelihood of $\sim 10$ standard deviations.
Attempts  to fit LASS data with this width also fail to give an
acceptable fit unless additional $s$-dependence is added to the
width; that leads to instablility in the mass range from
the $K\pi$ threshold to 825 MeV, where the LASS data begin.
If the scattering length is then fixed to the prediction of
Chiral Perturbation Theory, a width of at least 630 MeV is required.

\section {The phase of the $\kappa$ }
The phase variation of the $\kappa$ with mass will now be discussed.
It is well determined by BES data by two major
interferences: (a) with the channel $K_0(1430)K^*(890)$, (b) with
$K_1(1270)K$ and $K_1(1400)K$.

As a direct check that data are consistent with the phase variation
of eqn. (3), the $\kappa \to K\pi$ contribution to BES data may be
fitted in slices of $K\pi$ mass up to 1700 MeV; above that the $\kappa$
amplitude becomes too small to be determined reliably because of
uncertainty in $K_0(1950)$.
Four alternatives slice fits have been examined. In the most
restrictive, the phase is fitted in one bin at a time, keeping the
magnitude fixed to that of the global fit. All other amplitudes are
refitted freely in magnitude and phase except one; (one phase must be
fixed and the maximum likelihood method depends on one magnitude also
being fixed). Convergence is fastest if this is chosen to be the
largest amplitude, $[K_0(1430)K^*(890)]_{L=0}$ but alternatives
give the same result.
In the bin from 1400-1500 MeV, the phase cannot be determined
accurately because the $\kappa$ signal is swamped by the large
$K_0(1430)$ peak.

Fig. 4(a) shows results as points, compared with the global fit (full
line). There is no systematic tendency for points to move away from the
global fit. Note that in fitting BES data, the $K^*(890)\kappa$
amplitude has an overall phase fitted freely to the isobar model. The
curve on Fig. 4(a) has been drawn so that the phase goes to zero at the
$K\pi$ threshold; therefore only the phase variation with mass is
meaningful. Points are determined directly as deviations from the
global fit. The phase of the $[K^*(890)\kappa ]_{L=0}$ amplitude
relative to $[K^*(890)L_0(1430)]_{L=0}$ is $2.28^c \pm 0.02^c$ and for
$[K^*(890)\kappa ]_{L=2}$ is $4.82^c \pm 0.04^c$. These correspond to
the phases of elastic scattering of $\kappa$ from $K^*(890)$ at the
mass of the $J/\Psi$ and can in principle take any values.

\begin{figure}[htbp]
\begin{center}
\epsfig{file=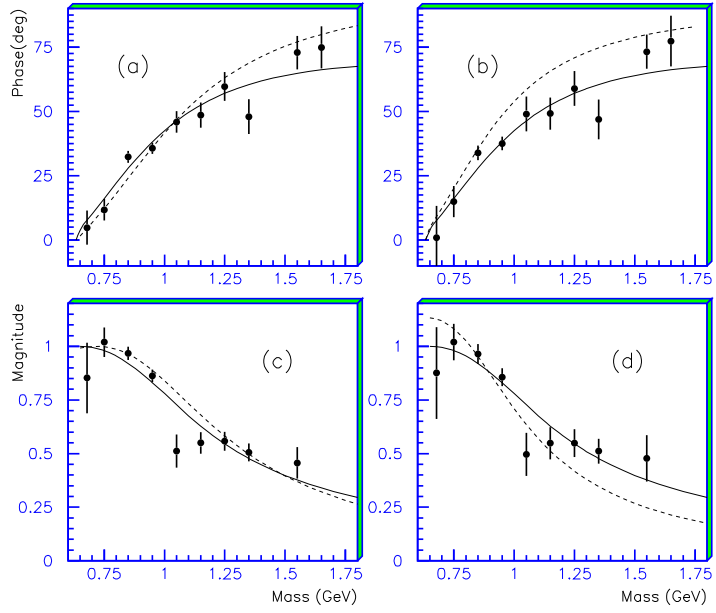,width=10cm}
  \caption[]{(a) Points show the phase of the $\kappa$
amplitude, determined bin-by-bin; the magnitude is fixed
from the global fit.
(b) phases when both magnitudes and phases are fitted in all
bins simultaneously.
(c) magnitudes when magnitudes and phases are both fitted
bin-by-bin.
(d) magnitudes when both magnitudes and phases are fitted
freely in all bins.
Full curves show the global fit.
Dashed curves in (a) and (c) show the optimum fit to BES
data alone. Dashed curves in (b) and (d) show the fit to
BES data with a Breit-Wigner amplitude of constant width.}
\end {center}
\end{figure}

The errors will now be discussed.
Statistics are very high (79K events) and it is common experience
elsewhere that, with these statistics, changes of log likelihood from
bin to bin may be above statistics; this can arise from the neglect of
small signals below 1\% intensity and also from approximations in the
Monte Carlo. Here it is found that statistical errors indeed need to be
multiplied by a factor 2.1 to allow for this.
Errors also include allowance for one standard deviation changes in
masses and widths of all resonances;
the main systematic error arises from parameters of $K_0(1430)$.

Next, the phase is fitted in all bins simultaneously, but fixing
the magnitude to the global fit.
Results are not shown because they are almost identical to Fig. 4(a).
The reason is that  separate bins contain different events.
The only correlation between bins arises from interference
with $K_1(1270)$ and $K_1(1400)$; one bin can pull the phase of
$K_1$ amplitudes slightly, and this reacts on other bins.
However, correlations between bins are only
$\sim 5\%$.

Finally, magnitude and phase are set free in all bins simultaneously.
It is however, necessary to fix the $\kappa$
amplitude in the bin from 1400 to 1500 MeV and above 1700 MeV;
the intensity in these bins is only 10\% of the integrated
$\kappa$ intensity.
Results are shown in Fig. 4(b).
There is no essential change from Fig. 4(a) but
errors increase because of statistical fluctuations in the
magnitude; this correlates with the fitted phase, because the
intensity in each bin depends on the real part of
interferences, i.e. on both magnitude and phase.

Fig. 4(c) shows the magnitude fitted in one bin at a time if both
magnitude and phase are set free together.
Fig. 4(d) shows the worst case, where magnitude and
phase are set free in all bins simultaneously; there is little
difference from Fig. 4(c).
Note that Figs. 4(c) and (d) show magnitudes; intensities
are the squares of these and are rather small for masses above
1200 MeV.

The conclusion from Figs. 4(a) and (b) is that the phase
variation of the $\kappa$ amplitude from BES data is the same
as the phase variation for elastic scattering within errors.
This is not a surprising result. It is well known that the
partial wave amplitude can be written in
the form $f(s) = N(s)/D(s)$ [24].
Here the numerator $N(s)$ is real
and arises from the left-hand cut, i.e. from driving forces. The
denominator $D(s)$ is complex and arises from the right-hand cut,
i.e. $K\pi$ rescattering. The pole terms in $D(s)$ are common to all
channels coupled to $K\pi$.
What the data tell us is that a single $\kappa$ pole term plus
$K_0(1430)$ fits the data up to $\sim 1700$ MeV.
This hypothesis can be tested directly by adding a further pole.
The LASS data rule out the possibility of a pole below 1950
MeV. Any further pole above that mass improves log likelihood
for BES data by less than three standard deviations.

The main sources of systematic error in this result
lie at masses above 1200 MeV. They arise from
(i) uncertainty in the precise line-shape for
$K_0(1430)$ and (ii) cross-talk with the small $K_0(1950)$
signal.
Fig. 5(a) shows the fit to LASS data.
There is a small systematic discrepancy in the mass range
around 1250 MeV. It is of similar magnitude to discrepancies
between Estabrooks et al. [25] and Aston et al. [8].
It can be reduced in several alternative ways.
One is to add a further factor $(1 + \beta s)$
to the phase of the $\kappa $ amplitude; the result is shown
in Fig. 5(b). The $\kappa$ pole moves to 697-i336 MeV. An
alternative is to allow some further inelasticity into $K_0(1430)$, due
to decays to $K^*\rho$ or $\kappa \sigma$; the LASS group quotes a $\pm
9\%$ normalisation uncertainty which allows a modest amount of such
inelasticity. A similar effect may be obtained by allowing some
inelasticity into the $\kappa$ amplitude; $\pi$ exchange between $K$
and $\pi$ allows production of the final state $\kappa \sigma$. A
calculation of the magnitude of such an effect would be valuable along
the lines suggested by Wu and Zou [26].

\begin{figure}[htbp]
\begin{center}
\epsfig{file=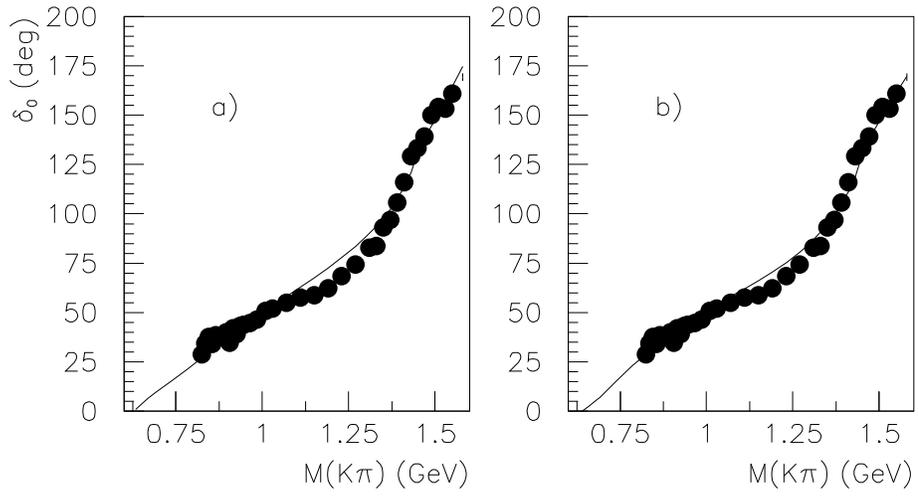,width=14cm}
\vskip -0.5cm
  \caption[]{(a) Optimum fit to LASS data from eqns. (1)--(4).
(b) improved fit allowing one extra parameter in fitting LASS
data.}
\end {center}
\end{figure}

Returning to Figs. 4(a) and (c), the dashed lines show the optimum fit
to the $\kappa$ for BES data alone. This requires $\alpha = 0$
in the exponential of equn. (2). The fit to BES data improves
by 3.8 standard deviations; the $\kappa$
becomes narrower, with a pole position of $753 - i319$ MeV.
However, the price is that the $K_0(1430)$ required
to fit LASS data is then definitely wider than that required by
BES data. This inconsistency suggests it is not a real physical
effect.

The FOCUS collaboration has recently measured the phase of the
$K\pi$ S-wave amplitude in $D^+$ decays to $K^-\pi ^+ \mu ^+ \nu$ [27].
The $\mu$ and $\nu$ are weakly interacting, so Watson's theorem
applies rigorously. They find a phase for the $K\pi$ S-wave in
excellent agreement with LASS data.
So BES, FOCUS and LASS data agree on the phase of the $K\pi$
S-wave, and hence with the $N/D$ formalism.
However, FOCUS find a phase in strong disagreement
with the phase of the $\kappa$ amplitude used by E791 for
the similar process $D^+ \to K^-\pi ^+ \pi ^+$.
The $\kappa $ amplitude needed by E791 is small; it is
important to check whether it can be fitted by the LASS
effective range formulae; if so, that would resolve
the discrepancy.

A final comment is that the phase of the $\kappa$ amplitude
could be affected by rescattering of its decay products
from the spectator pion. Such so-called triangle diagrams
have been discussed by Anisovich and Ansel'm [28]. This
rescattering is what generates the phase of the coupling
constant appearing in the isobar model. It could in
principle vary over the mass range of the $\kappa$. The
present analysis does not call for such an effect beyond
the combined experimental errors of BES and LASS data.
Another possible rescattering effect amongst final
state particles is the appearance of a t-channel pole
in $K\pi$ due to exchange of $K^*(890)$ or $\rho(770)$;
however such exchanges are the driving forces which
are conventionally believed to drive $K\pi$ elastic
scattering and hence the $\kappa$ pole.

\section {Alternative fits}
A second test has been made using for the $\kappa$
a Breit-Wigner amplitude of constant width, i.e. no
$\rho (s)$ factor in $\Gamma (s)$.
The best fit is shown by the dashed curves in Figs. 4(b) and (d).
This fit requires $M_1 = 718$ MeV, $\Gamma = 1045$ MeV,
corresponding to a pole position of M$_1 = 844-i444$ MeV.
The phase variation with mass is close to that of elastic
data. However, it requires a phase shift of $81^\circ$ at threshold
in elastic scattering, and this is clearly unphysical.

A fit to the $\kappa$ using $\Gamma = \Gamma _0\rho (s)$ gives a very
bad fit, worse than eqns. (1)--(3) by $>15$ standard deviations.
It requires $M_1 = 1035$ MeV, $\Gamma _0 = 725$ MeV, and a pole position
of $1043 -i410$ MeV. The problem is that it requires a phase
variation of $90^\circ$ between threshold and the resonance
mass; both BES and LASS data disagree strongly with this.
However, as the Ishida group has shown [29], the elastic data may be
fitted with a conventional Breit-Wigner resonance with
$\Gamma (s) \propto \rho(s)$ if one adds a
background phase decreasing rapidly with $s$, and cancelling a large
part of the resonant phase shift.
Both LASS and BES data can be fitted in this way, but only if the
background is the same in both sets of data within the errors.
This is algebraically equivalent to including
the $s$-dependence of the background into $\Gamma (s)$.
The fit reported by Komada [5] does in fact contain a background,
but details of this background are not given.
Komada reports a mass of $M= 882 \pm 24$ MeV. This is clearly
incompatible with the peak shown here in Fig. 2 unless such a
background is included and plays a major role.

\section {Conclusions}
The essential result from this analysis is
evidence for a low mass $K\pi$ S-wave enhancement
which definitely peaks close to threshold.
Eqns. (1)-(4) including the Adler zero give a
pole  at M = $(760 \pm 20(stat) \pm 40(syst)) - i(420
\pm 45(stat) \pm 60(syst))$ MeV.
The main systematic uncertainty lies in the width.
This is because the fitted magnitude above 1200 MeV is sensitive
to the large $K_0(1430)$ signal, hence
its precise line-shape.
The second result is that the fit reported here achieves
consistency with LASS data; the phase variation observed for the
$\kappa$ is consistent with that from elastic scattering. A similar
result was obtained earlier for the $\sigma$ [30].

From the best fit to both LASS and BES data, the $K\pi $ $I =0$
scattering length is $(0.23 \pm 0.04) m_\pi ^{-1}$. This is in close
agreement with the prediction of Chiral Perturbation Theory
at order $p^4$, namely $(0.19 \pm 0.02)
m_\pi ^{-1}$ [31].

\section {Acknowledgements}
   I wish to thank the BES collaboration for the use of the data
and the Royal Society for financial support of
collaboration between BEPC and Queen Mary, London, contract
Q771. I also wish to thank Dr. L.Y. Dong for extensive assistance
in processing data and Prof. B.S. Zou for discussions about the
physics over a period of many years.

\end{document}